# Coordinate System And Coordinate Transformations Based On Wave Nature Of Light

M. F. Yagan
Amman - Jordan

**Abstract**
The Classical Coordinate System is geometrical by nature with time being an external variable. Constructing a classical coordinate system employs a point-like signal with infinite speed. In Special Relativity Theory the speed is limited but the signal is a point-like particle (photon). If the oscillatory nature of light is considered, an event in absolute space is to be characterized by three coordinates namely, distance, time and phase. The Galilean transformation equations for space and time coordinates should be complemented by a third equation that accounts for the phase transformation. Wave equation remains invariant under such transformation and kinematical equivalence of inertial reference frames is conserved. Lorentz transforms apply to wave length and wave period of the exchanged light signal in a dynamic set-up.

**Classical Coordinate System (CCS)**
The construction of CCS explicitly adopts the concept of infinite speed of signal transmission. Starting from a certain point in absolute space (the origin of the CCS) the observer sends infinitely fast signals in all directions, the signal carrier is a point like object and event in this manner is determined by the occupation of a certain geometrical point in space at a given moment in time by this point like object. These point like objects reach all geometrical points in the space of the observer at the same instant leading to the evident result that any geometrical point (p) in the space has a time coordinate t equal to the time shown by the clock fixed to the origin o (t(p)=t(o)). The result is a pure geometrical coordinate system whose event points are correlated simultaneously and time is treated as an external variable. Therefore, the spatial distance between two event points in space represents the distance between the two simultaneous events taking place each at one of the points. At this point it's useful to remember that wave equation correlates two different events (defined as the phase of the propagating wave at two different points in space and time) causally, this definition of events leads to the obvious conclusion that transformations applied to wave equations should reflect the causality relations between events. This specific feature of wave equations will be the determining factor in choosing the coordinates transformations as shown below.

The CCS adapts perfectly with (Euclidian) geometry and classical kinematical analyses where time is externalized. Once a speed limit is imposed on signal transmission (in addition to the wave nature of the signal), the events in the system become causally related, the distances between events are no longer the distances between simultaneous events, leading necessarily to a chrono-geometrical construction of the coordinate system with time being an internal variable, this we call Causal Relational Coordinate System (CRCS).

**Causal Relational Coordinate System**

A (CRCS) is a system whose points represent events taking place in a specific geometrical location at a specific moment of time causally related to the event taking place in the origin O at a previous moment of time $t(O)$ as a consequence of the wave nature of the signal and the limit ness of it's speed.

Einstein's clock synchronization is based on the point like nature of light which reduces the definition of an event at a certain point in space and time to the simple occupation of that point by the light particle and obscures the causal relational aspect of a given coordinate system. Synchronization of clocks reduces the coordinate system back to CCS where all points have the same time coordinate as the origin regaining the simultaneity required for geometrical analysis.

In CRCS, space and time are correlated as coordinates by definition and have nothing to do with Einstein's space-time fabric, absolute space and absolute mathematical time are the background of all CRCSs, and coordinate transformations are precisely that, it's the coordinates that are transformed not the absolute space and time.

Consider two observers $O$, $O^1$ at the same point in absolute space and at rest relative to each other, each starts constructing his own CRCS by continuous emission of the same light signal with speed c, wave length $\lambda$ and wave period T, the result is two identical space and time charts of events. Each point in absolute space and time is defined in CRCS of observer O by event E and by an identical simultaneous event $E^1$ by observer $O^1$ in it's own CRCS. The events have the same geometrical coordinates measured from the common origin $x=x^1$, the same time coordinates $t=t^1$ measured from the same initial time at the common origin $t(O)=t^1(O^1)=0$ and the same phase $\Phi=\Phi^1$ relative to the identical initial phases at the common origin $\Phi(O)=\Phi^1(O^1)$.

Let the equation describing the causal relation between event E and the original event at $x=0$, $t=0$ be the following for observer O

$$\Phi(E)=f(x,t)/A= \cos(2\pi(x/\lambda - t/T)) \qquad (1)$$

where A is the amplitude, λ the wave length and T the period of the wave. Obviously the same equation applies for observer O¹ when it's proper variables are introduced into the equation.

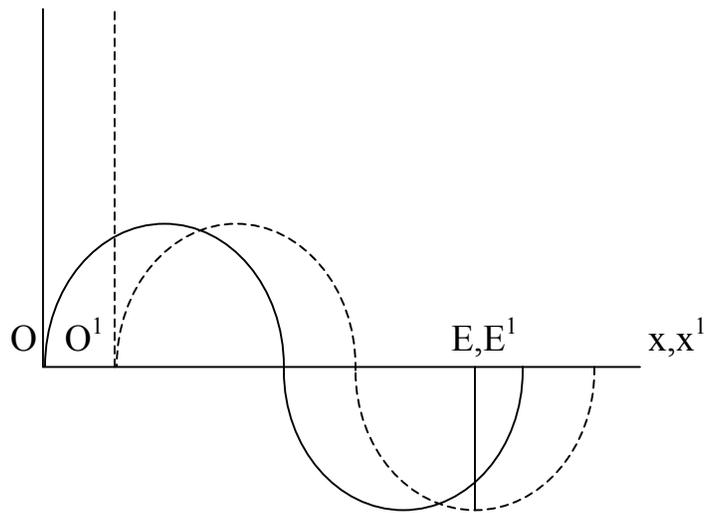

Figure 1

In the case where the two observers with their respective CRCS are moving relative to each other with velocity V along the common direction x, x¹ as shown in figure 1, the same point in absolute space and time will be defined by two absolutely simultaneous but non identical different events. Observer O defines the point by event E(x,t,Φ) while O¹ defines it by event E¹(x¹, t¹, Φ¹). Equation 1 above describes the causal relation between event E and the original event that took place at O(x=0,t=0, Φ(x=0,t=0)), proper transformation of the variables in the equation to the coordinate system of O¹ should keep the equation invariant. Obviously, any set of transformations used should transform equation 1 to describe the causal relation between event E¹ and the original event at origin O¹(x¹=0, t¹=0, Φ¹(x¹=0, t¹=0)) so three transformation equations are required for the three coordinates of events namely, position x, time t and phase Φ. The phase difference between the simultaneous events E and E¹ can be reconciled through the introduction of a phase factor K(t) in the cosine argument to transform the phase in E to the proper phase in E¹.

K(t) can be expressed as follows
$$K(t)= 2\pi Vt/cT \quad (2)$$
Consequently, the transformation equations required are the Galilean equations for kinematical (x,t) part of equation 1 and a third equation required for reconciling the causal (phase) part of the said equation

$$x= x^1 + Vt \quad (3a)$$
$$t= t^1 \quad (3b)$$
$$\cos(\theta)= \cos(\theta^1 - 2\pi Vt/cT) \quad (3c)$$
where $\theta=2\pi(x/\lambda - t/T)$ and
$\theta^1 =2\pi(x^1/\lambda - t^1/T)$
Applying these transforms to equation 1 we get
$$\Phi^1 (E^1)=f(x^1,t\,t^1)/A= \cos (2\pi(x^1/\lambda +Vt/\lambda - t^1/T) - 2\pi Vt/cT)$$
Introducing $\lambda= cT$, the transformed equation becomes
$$\Phi^1 (E^1)=f(x^1, t^1)/A= \cos (2\pi(x^1/\lambda - t^1/T) \quad (4)$$
The constancy of the absolute speed of light in absolute space is preserved at the same time Galilean transformations of coordinates prove to be valid for any type of wave equation as long as they are applied properly and in agreement with the causal context of wave equations. Lorentz transformations are the result of erroneous definition of events and the application of the CCS geometrical concepts.